\documentclass[iicol]{sn-jnl}

\usepackage{graphicx}%
\usepackage{multirow}%
\usepackage{amsmath,amssymb,amsfonts}%
\usepackage{amsthm}%
\usepackage{mathrsfs}%
\usepackage[title]{appendix}%
\usepackage{xcolor}%
\usepackage{textcomp}%
\usepackage{manyfoot}%
\usepackage{booktabs}%
\usepackage{hyperref}
\usepackage{algorithm}%
\usepackage{algorithmicx}%
\usepackage{algpseudocode}%
\usepackage{listings}%
\usepackage{hyperref}
\usepackage{gensymb}
\usepackage{textcomp}

\begin{document}

\title[Article Title]{\textbf{Beam Steering and Radiation Generation of Electrons in Bent Crystals in the Sub-GeV Domain}}

\author[1,2]{\fnm{R.} \sur{Negrello}} 
\author[1,2]{\fnm{M.} \sur{Romagnoni}} 
\author[2]{\fnm{A.} \sur{Sytov}} 
\author[1,2]{\fnm{N.} \sur{Canale}} 
\author[3,4]{\fnm{D.} \sur{De Salvador}} 
\author[1,2]{\fnm{P.} \sur{Fedeli}} 
\author[1,2]{\fnm{V.} \sur{Guidi}} 
\author[]{\fnm{V. V.} \sur{Haurylavets}} 
\author[5]{\fnm{P.} \sur{Klag}} 
\author[5]{\fnm{W.} \sur{Lauth}}
\author[2]{\fnm{L.} \sur{Malagutti}}
\author[1,2]{\fnm{A.} \sur{Mazzolari}} 
\author[2]{\fnm{G.} \sur{Patern\'o}}
\author[3,4]{\fnm{F.} \sur{Sgarbossa}} 
\author[6]{\fnm{M.} \sur{Soldani}} 
\author[]{\fnm{V. V.} \sur{Tikhomirov}}
\author[3,4]{\fnm{D.} \sur{Valzani}}
\author*[2]{\fnm{L.} \sur{Bandiera}}\email{bandiera@fe.infn.it}

\affil[1]{\orgdiv{Department of Physics and Earth Science}, \orgname{University of Ferrara}, \orgaddress{\street{Via Saragat 1}, \city{Ferrara}, \postcode{44122}, \state{Italy}}} 
\affil[2]{\orgdiv{INFN Ferrara}\orgname{}, \orgaddress{\street{Via Saragat 1}, \city{Ferrara}, \postcode{44122}, \state{Italy}}} 
\affil[3]{\orgdiv{Department of Physics}, \orgname{University of Padua}, \orgaddress{\street{Via Marzolo 8}, \city{Padua}, \postcode{35131}, \state{Italy}}} 
\affil[4]{\orgdiv{INFN Laboratori Nazionali di Legnaro}\orgname{}, \orgaddress{\street{Viale dell'Universit\'a 2}, \city{Legnaro}, \postcode{35020}, \state{Italy}}} 
\affil[5]{\orgdiv{Institut f\"ur Kernphysik}, \orgname{Universit\"at Mainz}, \orgaddress{\city{Mainz}, \postcode{55099}, \state{Germany}}}
\affil[6]{\orgdiv{CERN}, \orgname{European Organization for Nuclear Research}, \orgaddress{\city{Geneva 23}, \postcode{CH-1211}, \state{Switzerland}}}

\abstract{
We present an investigation into beam steering and radiation emission by sub-GeV electrons traversing a bent silicon crystal. Using 855, 600, and 300~MeV electron beams at the Mainz Microtron (MAMI), we explored orientational coherent effects in a 15-$\mu$m-thick silicon crystal bent along the (111) planes. Combined experimental and simulation analyses enabled the classification and quantitative assessment of the contributions from channeling, dechanneling, rechanneling, and volume capture to both beam deflection and radiation emission. Crystal steering remained effective even at 300~MeV, with measured channeling efficiencies exceeding 50\%, a record at such low energies. Channeling and volume reflection enhanced radiation emission by up to a factor of six compared to the misaligned orientation, highlighting strong orientational coherence effects in the sub-GeV regime. These findings confirm the feasibility of using bent crystals for efficient beam manipulation and high-intensity photon generation at low energies, supporting the development of novel light sources and beam control strategies at accelerator facilities operating in this energy range.}

\maketitle
\section{Introduction}\label{sec1}

Since the 1960s, it has been established that relativistic electrons and positrons traversing crystalline materials along specific lattice directions can generate radiation more intense than standard bremsstrahlung due to coherent interactions with the crystal lattice \cite{TerMikaelian, BaierKatkov, AkhiezerShulga}. This phenomenon, known as \textit{coherent bremsstrahlung}, was first investigated theoretically by Ferretti \cite{Fer}, Ter-Mikaelian \cite{TerMikaelian}, and Dyson and Überall \cite{Uber}, and was subsequently observed in 1960 at the Frascati laboratory by Diambrini-Palazzi and collaborators~\cite{bologna-1960}. Coherent bremsstrahlung arises when the momentum transfer to the medium satisfies the reciprocal-lattice condition, namely when it matches a reciprocal lattice vector. Its intensity, markedly enhanced with respect to ordinary bremsstrahlung, immediately suggested its potential as a compact source of intense X-rays and $\gamma$-rays \cite{BILOKON1983299}.

In 1963, the concept of channeling emerged from computer simulations of ion beams traversing properly oriented crystals \cite{robinson-1963}. Channeling occurs when the angle between the incoming charged particle's trajectory and a major crystallographic axis or plane is smaller than the so-called Lindhard angle, $\vartheta_L = \sqrt{2U_0/pv}$ \cite{Lin}, where $U_0$ is the depth of the axial or planar potential well, and $p$ and $v$ are the momentum and velocity of the incident particle, respectively. Under these conditions, correlations arise between successive collisions of the particle with atoms aligned along the same row or plane, allowing the discrete atomic potentials to be approximated by an average continuous potential \cite{Lin}. The potential well depth $U_0$ can vary significantly, typically ranging from tens of eV for planar orientations in silicon, to several hundred eV along crystallographic axes, and up to nearly 1 keV for high-$Z$ materials such as tungsten.

Channeled particles oscillate within the planar or axial potential wells, leading to a radiation emission process called \textit{channeling radiation} \cite{Kumakhov197617}. Although typically softer, this radiation is more intense than coherent bremsstrahlung. The intensity of channeling radiation is limited by the phenomenon of \textit{dechanneling} \cite{Beloshitsky, FlillerIII200547}, where particles leave the channeled state due to incoherent scattering with the crystal's electrons and nuclei.

Over the past decades, the phenomenon of channeling in bent crystals has been extensively investigated for its potential to steer charged particle beams at various accelerators, such as the SPS and the LHC at CERN~\cite{Scandale:2010:SPS, scandale-2007}, at Fermilab~\cite{PhysRevSTAB.5.043501} and the U-70 synchrotron~\cite{Afonin:2001}. Originally proposed by Tsyganov in 1976~\cite{Tsyganov:1976:TM-682,Tsyganov:1976:TM-684}, this technique exploits planar channeling within bent atomic planes to guide particles along predetermined trajectories by confining them within potential wells shaped by the crystal's curvature. A distinctive feature of bent crystals is the ability to deflect overbarrier particles via volume reflection \cite{taratin-1987}, in which the transverse momentum is reversed by the interaction with the effective potential barrier of a bent atomic plane. Since volume reflection affects particles not subject to dechanneling, it achieves higher deflection efficiencies, although with smaller deflection angles compared to channeling, for both positively \cite{PRL.QM, Bagli2014effVsRad} and negatively charged particles \cite{scandaleQM}.

Channeling and volume reflection in bent crystals have been investigated for electron energies ranging from a few~GeV at SLAC and CERN up to the hundred-GeV scale at CERN, including studies aimed at radiation generation~\cite{scandale-2009, bandiera-2013, bandiera-Nim-2013, nielsen-2019}.
In the sub-GeV region, only a limited number of studies exist, mostly focused on 855~MeV electrons at MAMI~\cite{PhysRevLett.112.135503, sytov-2017, bandiera-2015, bandiera-2021, haurylavets-2021, korol-2021-epjd}. A further study has addressed beam steering of 255~MeV electrons at SAGA~\cite{TAKABAYASHI2018153}, using a crystal not optimized for high-efficiency deflection.
To date, radiation-emission measurements have not been reported for electron energies below 855~MeV.

In this study, we investigate the behavior of sub-GeV electrons in bent silicon crystals, with a focus on beam steering and radiation generation as a function of beam energy. This research combines experimental measurements, simulations, and detailed data analysis to provide a comprehensive understanding of the underlying processes. Specifically, this study extends the investigation of radiation emission in bent crystals down to 300 MeV. 
This energy range is particularly attractive for the development of innovative radiation sources, as it can enable intense hard X-ray (10 keV$-$1 MeV) and $\gamma$-ray (1$-$100 MeV) emission in compact accelerator and therefore it is of significant interest for radiation-source concepts at several LINAC facilities worldwide~\cite{HEPS,SLAC,CLEAR,SPARCLAB,ARES2024}, while also providing advanced capabilities for beam steering. Exploiting coherent effects in bent crystals at these energies offers a promising, efficient and cost-effective way to produce light sources for X-ray microscopy and industrial radiography. Moreover, sub-GeV beams facilitate the use of high-$Z$ materials and large-emittance beams, potentially enhancing crystalline radiation production. 

The paper is organized as follows. Section~\ref{sec:channeling-bent} introduces the theoretical framework of channeling in bent crystals. Section~\ref{setup} describes the experimental setup and the crystal sample. Section~\ref{sec-simmethod} describes the simulation methodology. Section~\ref{sec2-Defl} presents the beam-steering measurements, including the deflection efficiency and dechanneling effects, and provides a critical comparison with Monte Carlo simulations. Section~\ref{subsecrad} is devoted to the analysis of radiation emission, including a spectral decomposition based on particle dynamics obtained from dedicated simulations. Finally, the results are discussed in section \ref{sec5-discussion}, where possible applications are also highlighted.

\section{Channeling in Bent Crystals}
\label{sec:channeling-bent}

\begin{figure*}[h!]
    \centering
    \includegraphics[width=0.9\linewidth]{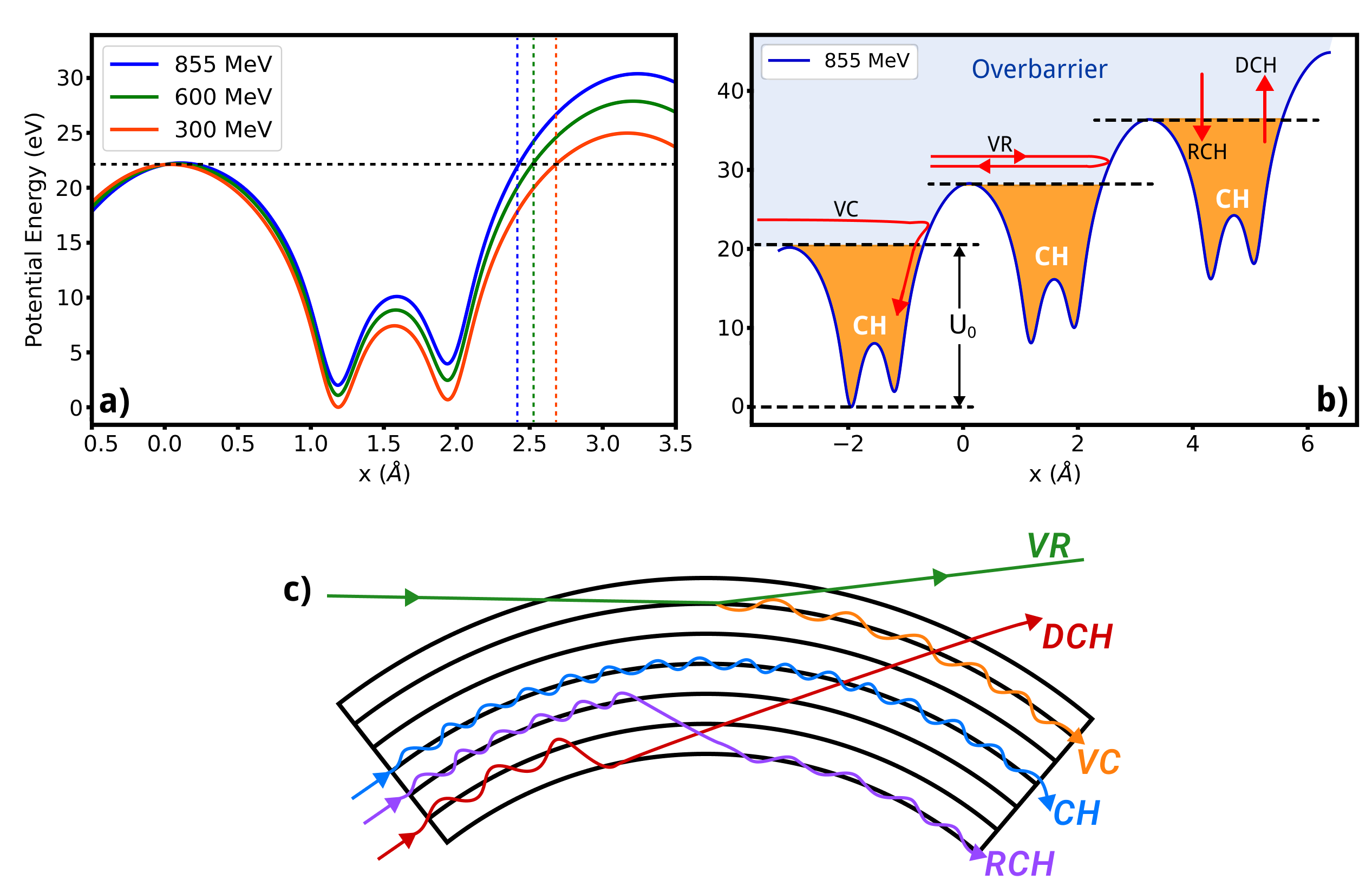}
    \caption{(a) Effective potential energy for electrons channeling in bent Si(111) planes. The curves correspond to beam energies of $300~\text{MeV}$ (red), $600~\text{MeV}$ (green), and $855~\text{MeV}$ (blue).  
    (b) Same potential for 855 MeV, with shaded regions indicating the potential wells where electrons are channeled (CH), where arrows show the other processes occurring in bent crystals: dechanneling (DCH), rechanneling (RCH), volume capture (VC) and volume reflection (VR).  
    (c) Schematic view of CH, RCH, DCH, VC and VR trajectories in a bent crystal.}
    \label{fig:pot&bent}
\end{figure*}

This section introduces channeling and associated coherent effects in bent crystals. To accurately describe the dynamics of sub-GeV electrons, it is essential to account for the modification of the planar potential induced by the crystal curvature.
In a bent crystal, the curvature of the atomic planes acts as a guiding structure, forcing channeled particles to follow the bending. This introduces a non-inertial centrifugal force in the co-rotating reference frame of the particle, which modifies the interaction dynamics.  
This effect can be described by an effective potential 
\begin{equation}
U_{eff}(x) = U(x) - \frac{pv}{R} x 
\end{equation}where $U(x)$ is the continuous interplanar potential, $p$ is the particle momentum and $R$ is the bending radius \cite{biryukov-1997}. For Si(111), $U(x)$ presents a ``double-dip'' shape due to the alternating interplanar distances $d_{pl}^L=2.35$~\AA{} and $d_{pl}^S=0.78$~\AA{} \cite{biryukov-1997}.
Figure~\ref{fig:pot&bent}(a) shows the variation of $U_{eff}(x)$ as a function of particle momentum for the cases investigated in this paper: 300, 600 and 855 MeV electrons channeled in Si(111) planes with $R \approx 32.6$ mm. The potential has been calculated using the commonly employed Doyle-Turner approximation~\cite{doyle-1968,biryukov-1997}. A black horizontal line at the potential value corresponding to $x=0$ is drawn to highlight the shape of the potential well, felt by the electron, for the three energy considered. It is evident how both the depth and the width of the potential well increase as the beam energy decreases.
The modification of the potential well is governed by the ratio $R/R_C$, where $R_C = pv/E_{0}$ is the critical radius and corresponds to the curvature limit below which the potential well disappears, forbidding channeling \cite{biryukov-1997}. Here, $E_{0}$ represents the maximum interplanar electric field. 

Consequently, bending the crystal reduces the channeling critical angle ($\vartheta_L^b$) compared to the straight case ($\vartheta_L$). It can be computed as: \begin{equation}
\vartheta_L^b = \vartheta_L \left( 1 - \frac{R_C}{R} \right).
\end{equation}

If the angle between the particle trajectory and the atomic planes ($\vartheta_x$) is smaller than $\vartheta_{L}^b$, the particle may be captured into a bound state. This condition is necessary but not sufficient for channeling: depending on the impact parameter with respect to the plane, a particle can still have a transverse energy $E_T$ exceeding the depth of the potential well, even when $\vartheta_x < \vartheta_{L}^b$.

A particle is considered in a \textit{bound state} when its transverse energy $E_T$, defined as:
\begin{equation}\label{eq:ET}E_T = \frac{pv}{2}\vartheta_x^2 + U_{eff}(x)\end{equation}is lower than the potential barrier maxima $U_{0}$. Conversely, it is in an \textit{unbound state} (or overbarrier) if $E_T > U_{0}$. 

Figures~\ref{fig:pot&bent}(b) and (c) illustrate the main classes of electron motion in the field of bent atomic planes:
\begin{itemize}
\item \textbf{Channeling (CH):} electrons with $E_T<U_{0}$ are trapped in the planar potential well, oscillate around planes, and follow the crystal curvature. 
\item \textbf{Dechanneling (DCH):} incoherent scattering increases $E_T$ until the channeled particle overcomes the barrier and leaves the bound state. 
\item \textbf{Rechanneling (RCH):} occurs when previously dechanneled particles are recaptured in the channeling state, due to the loss of transverse energy caused by incoherent scattering.
\item \textbf{Volume Reflection (VR):} consists of the reversal of the transverse momentum of overbarrier particles whose trajectories become locally tangent to bent planes within the crystal, producing a net angular deflection opposite to the crystal bending (typically of the order of $\vartheta_{L}$). It can occur for particles with incidence angle $\vartheta_{L}^b<\vartheta_x<\vartheta_b$, where $\vartheta_b$ is the crystal bending angle.
\item \textbf{Volume Capture (VC):} occasionally, an overbarrier particle reaching the tangency region may, instead of being volume reflected, lose sufficient transverse energy through scattering to become trapped in the potential well, thus transitioning directly into the channeling state.
\end{itemize}

The primary mechanism limiting deflection efficiency is dechanneling. This process is characterized by the \textit{dechanneling length} $L_d$, defined as the average distance over which particles maintain the channeling regime before being scattered out of the potential well. $L_d$ is conventionally defined by assuming an exponential decay of the channeled particle fraction $N_{ch}$ as a function of the penetration depth $z$:
\begin{equation}\label{eq:negexp}
    N_{ch}(z)=A_0 \exp(-z/L_d)
\end{equation}
where $A_0$ is a normalization factor. This is a standard diffusion approach and is well grounded for describing dechanneling induced by scattering with valence electrons for non-relativistic ions and positively charged particles \cite{Beloshitsky, biryukov-1997}. However, for negatively charged particles, which oscillate around the crystal planes, it is less accurate as it is dominated by Coulomb scattering with nuclei and core electrons that induce frequent strong perturbations in the transverse motion \cite{tikhomirov-2017}. Nevertheless, $L_d$ remains a crucial parameter to qualitatively characterize the stability of the channeling regime, making its evaluation essential for applications.

\section{Experimental Setup}\label{setup}

\begin{figure*}
    \centering
     \includegraphics[width=1\linewidth]{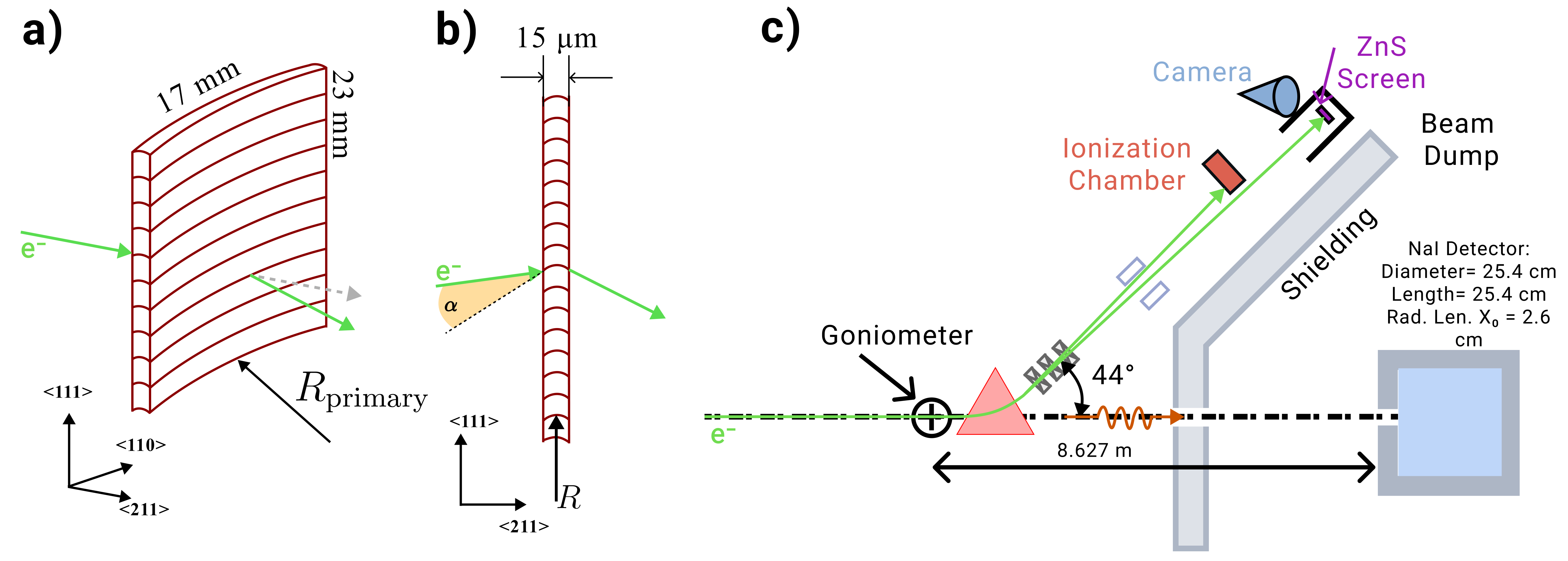}
    \caption{(a) Three-dimensional view of the bent Si(111) crystal mounted on the holder that imparts the mechanical deformation via the primary curvature $R_{primary}$. (b) Side view showing the quasi-mosaic curvature radius $R$ of the (111) planes and the beam incidence angle $\alpha$. (c) Scheme of the experimental setup at MAMI B, adapted from \cite{bandiera-2021}.}
    \label{fig:setup}
\end{figure*}
We conducted an experiment at the Mainzer Mikrotron (MAMI) in Mainz (Germany).
We employed a silicon crystal specifically optimized for the steering of sub-GeV electrons by maximizing the deflection efficiency via channeling. This was achieved by using a plate-like sample with a reduced thickness along the beam direction to minimize incoherent scattering and to be of the order of the dechanneling length in the energy range investigated (e.g., $L_d \simeq 21~\mu$m at 855 MeV \cite{PhysRevLett.112.135503}) so that the majority of channeled electrons would remain confined throughout the crystal. As illustrated in Figure~\ref{fig:setup}(a), the crystal features transverse dimensions of $23 \times 17~\mathrm{mm}^2$ and a thickness of $(15 \pm 1)~\mu\mathrm{m}$, with crystallographic orientations selected to induce bending of the (111) planes, parallel to the 15 $\mu$m $\times$ 23 mm face.
To achieve the desired curvature, the sample was mounted on a mechanical holder imposing a primary curvature, $R_{primary}$, along the $\langle 211 \rangle$ crystal axis. Exploiting the elastic anisotropy of silicon, this deformation induces a secondary curvature, $R$, along the $\langle 111 \rangle$ direction, due to the \textit{quasi-mosaic effect}~\cite{Camattari:ks5467, germogli-2015, IvanovQM} resulting in the bending of the (111) planes responsible for the electron beam steering (see Figure~\ref{fig:setup}(b)). The crystal was designed to have a bending angle of the (111) planes, $\vartheta_b \approx 460~\mu$rad, corresponding to a curvature radius $R \simeq \ell / \vartheta_b \approx 32.6$ mm for a crystal length of $15\,~\mu\mathrm{m}$. In Figure~\ref{fig:setup}(b), $\alpha$ defines the goniometer angle, that is the angle between the incoming particle beam direction and the tangent to the atomic bent planes at the entrance of the crystal.

Finally, the experimental bending radius is significantly larger than the critical radius $R_C$ for all investigated energies. Specifically, $R_C$ is estimated to be $1.53$~mm at 855~MeV, down to $1.07$~mm and $0.54$~mm for 600~MeV and 300~MeV, respectively.

The experiment was carried out in the Hall B of MAMI with 855, 600 and 300 MeV electrons. The experimental setup  is depicted in Figure~\ref{fig:setup}(c) and is the same as in \cite{bandiera-2021}. Electrons traversed the crystal, were deflected by a dipole magnet, and their position was measured on a LYSO screen positioned 6020 mm downstream of the center of crystal, to measure the beam profile downstream of the crystal and calculate its deflection. The LYSO screen was 200 $\mu$m thick and a camera was used to collect the emitted light. Precision in crystal alignment was achieved through a combination of a high-precision goniometer with five degrees of freedom (3 rotational and 2 translational) and analysis of the beam's deflection patterns. Beam focusing was performed using quadrupole lenses, achieving a beam size of approximately  100 $\mu$m and beam divergence of about 30 $\mu$rad. The divergence is well below the Lindhard critical angle (a few hundreds of $\mu$rad) at the considered energies. The emitted radiation obtained by the interaction of the $e^{-}$ beam with the bent crystal was measured by a scintillator detector. It consisted of a 25.4 cm in both diameter and length (corresponding to 10 $X_0$) thallium doped sodium iodide—NaI(Tl)—crystal read out by three photomultiplier tubes. The detector was positioned behind a collimator that only accepted photons within a 40 mm aperture located 8627 mm from the crystal \cite{bandiera-2015}, ensuring a collimation angle of 2.32 mrad, corresponding to about $3.9$, $2.7$, and $1.4$ times the relativistic emission cone ($1/\gamma$) for 855, 600, and 300 MeV, respectively.
Calibration of the NaI  detector was performed using the natural radioactive isotopes $^{40}$K (1.4161 MeV) and $^{208}$Tl-$^{228}$Th (2.614 MeV) \cite{bandiera-2015, bandiera-2021}.

\section{Simulation Methods}\label{sec-simmethod}

To interpret the experimental results and gain insight into the microscopic dynamics of electrons inside the bent crystal, we performed high-statistics Monte Carlo simulations.
Over the years, several approaches have been developed to describe the motion of relativistic charged particles in crystalline media, including methods based on the numerical solution of the classical equations of motion in the continuous crystal potential~\cite{sytov2015crystal} as well as molecular dynamics techniques~\cite{solovyov2012}.

Here, the propagation of relativistic electrons through the oriented silicon crystal was simulated using the \texttt{G4ChannelingFastSimModel}~\cite{Sytov-2024}, a dedicated physics model based on the algorithm developed in~\cite{guidi2012radiation} implemented within the \textsc{Geant4} simulation toolkit~\cite{Agostinelli2003, Allison2016}.
This model describes particle transport by solving the classical equations of motion for charged particles in the continuous potential of atomic planes. The simulation integrates the equations of motion using a Runge-Kutta algorithm, simultaneously accounting for incoherent scattering processes.  To ensure a precise sampling of the channeling oscillations  (of the order of $\approx \mu\text{m}$) and incoherent scattering, the integration step length was effectively maintained in the range of  $50$--$100$~nm.  These segments are also classified based on the transverse energy $E_T$ relative to the potential barrier $U_0$ in \textit{bound} and \textit{unbound}. This methodology also enables the decomposition of the deflected-particle distribution into distinct contributions associated with the different beam-crystal interaction mechanisms.

For the calculation of the emitted radiation, the simulation employs the Baier-Katkov quasiclassical formalism~\cite{guidi2012radiation,bandiera2015radcharm++,sytovPRAB2019}, which evaluates the radiation spectrum by integrating along the particle trajectory through the crystal. This approach enables the correlation of the particle dynamics with the emitted radiation, thereby quantifying and highlighting the contributions of the various processes occurring in bent crystals (channeling, dechanneling, rechanneling, and overbarrier motion) to the total radiation spectrum. 

\section{Beam Steering}\label{sec2-Defl}

\begin{figure*}[h!]
    \centering
    \includegraphics[width=0.9\linewidth]{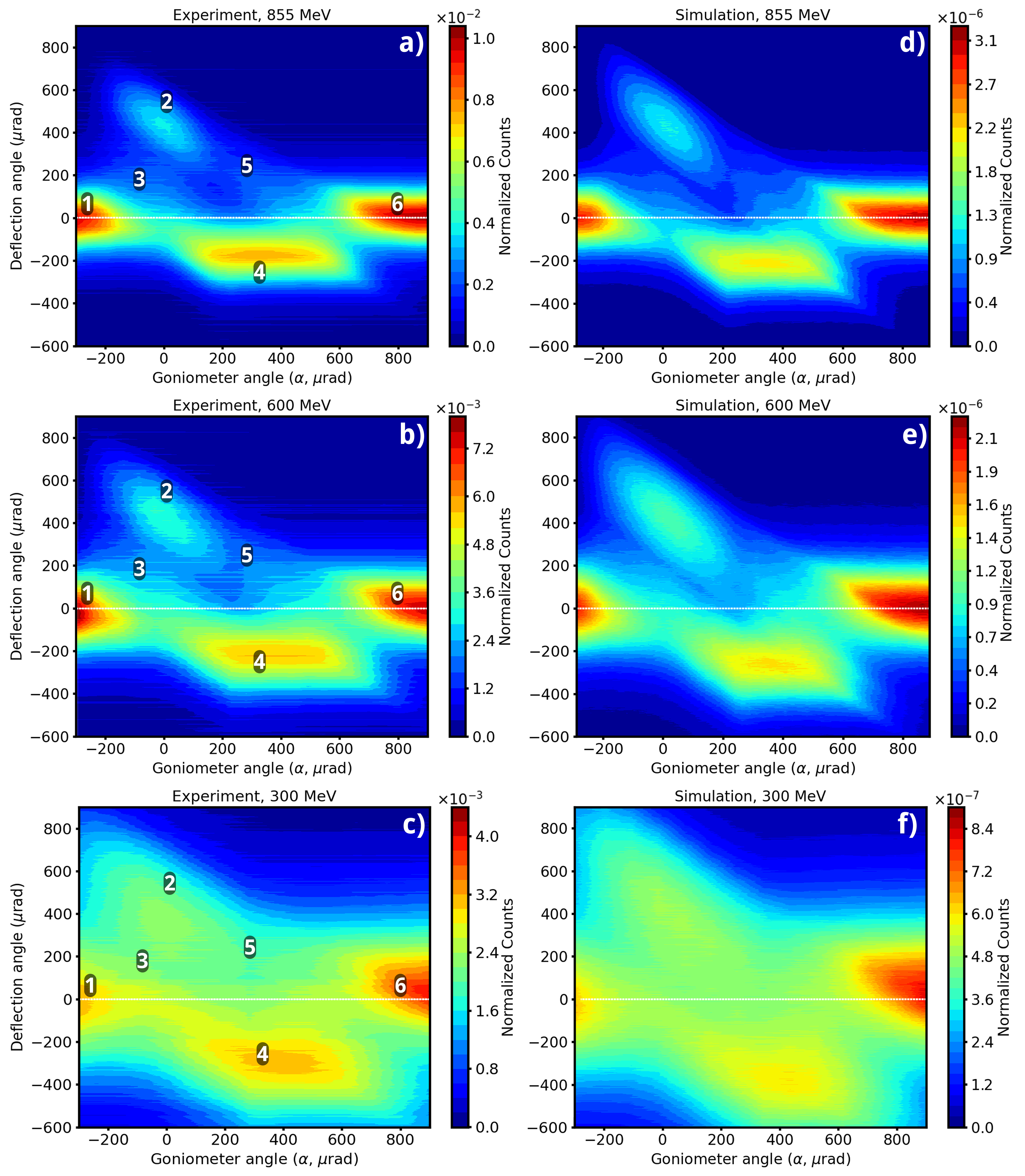}
    \caption{Experimental (a–c) and simulated (d–f) angular scans, showing the deflection angle vs. the goniometer angle $\alpha$ at the investigated energies. The distinct regions correspond to: (1) and (6) amorphous orientation, (2) channeling, (3) dechanneling, (4) volume reflection, and (5) volume capture.}
    \label{fig:scan}
\end{figure*}

\begin{figure}[h!]
    \centering
    \includegraphics[width=0.95\linewidth]{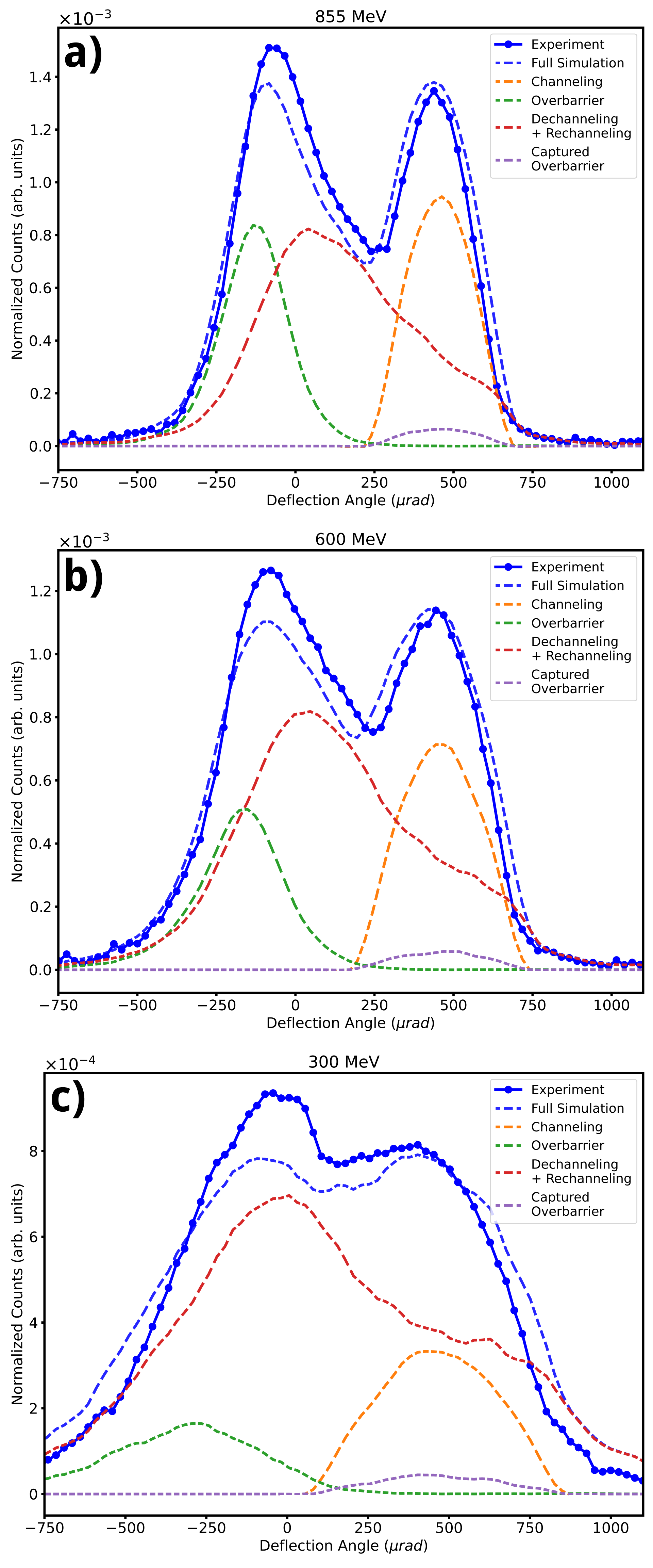}
    \caption{Normalized angular distributions of the deflected beam under channeling conditions for (a) 855 MeV, (b) 600 MeV, and (c) 300 MeV. Blue solid lines with markers represent experimental data, while dashed blue lines show the total simulated distributions. Simulated sub-distributions identify specific dynamical processes: channeling (orange), dechanneling+rechanneling (red), overbarrier (green), and captured overbarrier (purple).}
    \label{fig:defltag}
\end{figure}

In this section, we report the measurements of deflection as a function of energy and compare the results with dedicated Monte Carlo simulations using the methodology described in Sec. \ref{sec-simmethod}. This study presents the first investigation of electron beam steering in bent crystals as a function of energy in the sub-GeV range.

Initially, we recorded the angular distribution of particle deflections after interaction with the crystal versus the beam incidence angle $\alpha$ with respect to the atomic (111) planes. The reference orientation $\alpha \approx 0$ was experimentally defined as the crystal position that yields the maximum channeling efficiency in the angular scans. Figure~\ref{fig:scan} presents the comparison between the measured (panels a–c) and simulated (panels d–f) angular scans for the three investigated energies. The experimental color scale represents normalized counts proportional to the LYSO detector signal, while the simulated color scale is normalized to the total number of simulated events.
Six characteristic regions are clearly identifiable: (1) and (6) amorphous orientations, (2) channeling deflection, (3) dechanneling, (4) volume reflection, and (5) volume capture. The simulations closely reproduce the experimental patterns across all energies, confirming the accuracy of the model.

The profiles of the deflected particles, aligned for optimal channeling ($\alpha \approx 0$) at various energies, are illustrated in Figure~\ref{fig:defltag}. Experimental data are represented by solid blue lines with markers, whereas simulations are depicted by dashed blue lines. The deflection distribution exhibits two primary peaks. The peak on the right corresponds to channeling, centered around the nominal bending angle. The left peak corresponds to particles that are overbarrier at the crystal entrance, including those that are deflected toward the side opposite to the crystal bending  direction\cite{tarantin-1979}.
The comparison reveals good agreement at 855~MeV and 600~MeV, with good reproduction of the width of the deflected beam distribution and a small difference in the relative heights of the channeling and overbarrier peaks. At 300 MeV, the simulations are less effective in reproducing the experimental deflection profile, as the width of the deflected distribution and the relative height of the two peaks are not fully matched. The differences are likely due to residual crystal torsion, i.e., a parasitic twist of the crystallographic planes around the beam axis induced by the mechanical bending, which causes the optimal channeling angle to vary across the transverse beam spot, effectively reducing the measured efficiency~\cite{Mazzolari2021}.
The profiles clearly show that, as the energy decreases, the broadening of the peaks becomes more pronounced; this occurs for different reasons. Specifically, the increased multiple scattering at lower energies spreads the particles over a wider angular range, enlarging the overbarrier peak. On the other hand, the broadening of the channeling peak is influenced by the critical angle, $\vartheta_L^b$, which increases as the energy decreases (see Tab \ref{tab:eta}). This expands the angular acceptance, allowing more particles to enter the channeling regime. These effects are particularly evident at 300 MeV, where both scattering and larger critical angle contribute to a much broader distribution, making the peaks more challenging to distinguish.

\begin{table}[ht]
\centering

\fontsize{11}{14}\selectfont

\setlength{\tabcolsep}{8pt} 

\renewcommand{\arraystretch}{1.} 

\caption{Experimental and simulated channeling deflection efficiencies ($\eta$) for the different energies and therefore at different $R/R_C$.}
\label{tab:eta}

\begin{tabular}{@{}lccc@{}}
\toprule
Parameter & \multicolumn{3}{c}{Energy [MeV]} \\
\cmidrule(l){2-4} 
 & 855 & 600 & 300 \\
\midrule
$R/R_C$ & 21 & 31 & 60 \\
$\vartheta_L^b$ $[\mu\text{rad}]$ & 212 & 257 & 370 \\
$\eta_{exp} [\%]$ & 39 $\pm$ 2 & 45 $\pm$ 4 & 54 $\pm$ 5 \\
$\eta_{sim} [\%]$ & 44 $\pm$ 2 & 49 $\pm$ 3 & 61 $\pm$ 4 \\
\bottomrule
\end{tabular}

\end{table}

These factors have direct consequences for channeling efficiency $\eta$, which is derived from the deflection profiles. The channeling efficiency is defined as the ratio of the number of deflected particles to the total number of events in the deflection distribution. $\eta$ can be estimated as the integral within  $3\sigma$ of the half-Gaussian fitted to the deflected peak, multiplied by two and divided by the integral of the total curve. The values of $\eta$ were derived for the different energies for both experimental and simulated distribution (see Tab.\ref{tab:eta}). These values are consistent with those reported in previous experiments at 855 MeV \cite{sytov-2017, PhysRevLett.112.135503}. Interestingly, despite the enhanced multiple scattering and the expected increase in the dechanneling contribution at lower energies, the channeling efficiency does not decrease. On the contrary, it rises from 39\% to 54\% as the particle energy decreases. This seemingly counterintuitive behavior can be explained by the increase in the ratio $R/R_C$ at lower energies, which results in a broadening and deepening of the effective potential well (see Figure~\ref{fig:pot&bent}(a)). In particular, the larger potential well at lower energies results in a larger phase space for particles to enter and remain in the channeling regime. The channeling efficiency exceeding 50\% at 300~MeV represents more than a fivefold increase over previous results at comparable energies ($\approx 10\%$ in \cite{TAKABAYASHI2018153}) and is highly relevant for applications in sub-GeV accelerators. It is worth noting that the simulated deflection efficiency follows the experimental trend; nevertheless, the simulated values are higher by a few percent compared to the experimental ones. This can be attributed to crystal torsion, as described above.

Since the simulations reproduce the experimental results quite well, we exploited them to separate the distinct contributions of the various electron dynamics processes to the total deflection distributions. This was achieved by monitoring the evolution of the particle's $E_T$ relative to $U_0$ along the entire trajectory. This approach allowed us to tag each electron based on its history of bound and unbound states, effectively distinguishing stable channeling from dechanneling, rechanneling, and overbarrier motion. Figure~\ref{fig:defltag} also shows such simulated sub-distribution for channeled, overbarrier, dechanneled+rechanneled and captured overbarrier electrons (colored dashed lines). By analyzing these components, it is possible to identify the contributions shaping the overall deflection profile. As the beam energy decreases from 855~MeV to 300~MeV, several trends emerge. First, the width of the channeling distribution (orange dashed lines) broadens, reflecting the increase in $\vartheta_L^b$ at lower energies. Second, the fraction of purely overbarrier particles (green dashed lines) decreases, as a larger fraction of the beam fulfills the channeling condition at the crystal entrance for reduced energies. This leads to an enhanced contribution from particles undergoing at least one dechanneling event (red dashed lines). In contrast, the contribution from over-barrier captured particles (purple dashed lines) remains relatively constant across the explored energy range. At 300~MeV, the build-up of the deflected peak is predominantly driven by rechanneling processes rather than by particles remaining in the initial channeling state.  This highlights the crucial role of dechanneled and rechanneled particles in contributing to the deflected peak.

Finally, one may also notice that the peak position of channeling in the full deflection distribution (blue curves in Figure~\ref{fig:defltag}, either experimental or simulated) is about $(445 \pm 5)~\text{\textmu rad}$, which is shifted downward by about 15~\textmu rad with respect to the peak position of the simulated channeled component alone (orange curve in Figure~\ref{fig:defltag}), located at $(460 \pm 5)~\text{\textmu rad}$, corresponding to the curvature radius $R = (32.6 \pm 0.4)$ mm. Indeed, only the latter represents the actual bending angle of the crystal, consistent with the value measured by X-ray diffraction and used as an input for the simulation. One may therefore conclude that in the full deflection distribution, the channeling peak position is slightly reduced by the convolution with the dechanneled electrons' distribution, and that the measured channeling peak does not correspond exactly to the bending angle.

\subsection*{Dechanneling}
\begin{table*}[ht]
\centering
\caption{Summary of dechanneling lengths at the three beam energies. $L_{d,\text{exp fit}}$ and $L_{d,\text{sim fit}}$ are obtained from angular fits, and $L_{d,\text{with rech}}$ and $L_{d,\text{without rech}}$ from the exponential decay including or excluding rechanneling.}
\label{tab:L_dech}
\begin{tabular}{@{}p{1.5cm} p{2cm} p{2.cm} p{2.cm} p{2.cm} p{2.cm}@{}}
\toprule
Energy [MeV] 
& $L_{d,\text{exp fit}}$ [$\mu$m] 
& $L_{d,\text{sim fit}}$ [$\mu$m] 
& $L_{d,\text{with rech}}$ [$\mu$m] 
& $L_{d,\text{without rech}}$ [$\mu$m] \\
\midrule
855  & $20 \pm 2$ & $21 \pm 2$ & $21 \pm 1$ & $16 \pm 1$ \\
600  & $17 \pm 2$ & $16 \pm 3$ & $18 \pm 1$ & $13 \pm 1$ \\
300  & -- & -- & $15 \pm 1$ & $9 \pm 1$ \\
\bottomrule
\end{tabular}
\end{table*}

The deflection profiles are also used to determine the dechanneling length, $L_d$, which is a crucial parameter for beam steering capabilities. To extract this parameter, we fitted the deflection distribution using a function that accounts for the dynamics in bent crystals, following the approach detailed in Refs.~\cite{sytov-2017, wistisen-2016, bagli-2017}. The normalized deflection profiles are fitted with a global function of the deflection angle $\vartheta_{defl}$, $(1/N)dN/d\vartheta_{defl}$, expressed as the sum of three distinct components~\cite{sytov-2017}:

\begin{equation}
\frac{1}{N}\frac{dN}{d\vartheta_{defl}} = \frac{d f_{ch}}{d \vartheta_{defl}} + B_{ob}\frac{d f_{ob}}{d \vartheta_{defl}} + \frac{d f_{dech}}{d \vartheta_{defl}},
\label{eq:total}
\end{equation}

\noindent where the terms account for channeling ($f_{ch}$), overbarrier ($f_{ob}$), and dechanneling ($f_{dech}$), respectively. The coefficient $B_{ob}$ represents the relative weight of the overbarrier component.

The channeling peak is modeled by a Gaussian~\cite{sytov-2017}:
\begin{equation}
\frac{d f_{ch}}{d \vartheta_{defl}} = \frac{A_{ch}}{\sigma_{ch} \sqrt{2\pi}} \exp\!\left[-\frac{(\vartheta_{defl} - \vartheta_{ch})^2}{2\sigma_{ch}^2}\right],
\label{eq:ch}
\end{equation}
with amplitude \(A_{ch}\), mean deflection angle \(\vartheta_{ch}\), and standard deviation \(\sigma_{ch}\).

The overbarrier component is described as a sum of two Gaussians, with common centroid \(\vartheta_{ob}\) and width \(\sigma_{ob}\)~\cite{sytov-2017}:
\begin{align}
\frac{d f_{ob}}{d \vartheta_{defl}} &= \frac{A_{ob}}{\sigma_{ob} \sqrt{2\pi}} \exp\!\left[-\frac{(\vartheta_{defl} - \vartheta_{ob})^2}{2\sigma_{ob}^2}\right] \notag \\
&\quad + \frac{1-A_{ob}}{r\,\sigma_{ob}\sqrt{2\pi}} \exp\!\left[-\frac{(\vartheta_{defl} - \vartheta_{ob})^2}{2 r^2 \sigma_{ob}^2}\right],
\label{eq:hr}
\end{align}
where \(A_{ob}\) is a relative weight and \(r\) defines the width ratio between the two Gaussian terms.

The dechanneling contribution can be approximated as a convolution of an exponential distribution with the first gaussian in Eq. \ref{eq:hr}, as in~\cite{sytov-2017}:
\begin{equation}
\begin{aligned}
\frac{d f_{dech}}{d\vartheta_{defl}} &= \frac{A_{dech}}{2\,\vartheta_{dech}} \exp\!\left[ \frac{\sigma_{ob}^{2}}{2\,\vartheta_{dech}^{2}} +\frac{\vartheta_{ch}-\vartheta_{defl}}{\vartheta_{dech}} \right] \times \\
&\quad \Bigg\{ \operatorname{erf}\!\left( \frac{\vartheta_{ob}-\vartheta_{defl} + \sigma_{ob}^{2}/\vartheta_{dech}}{\sqrt{2}\,\sigma_{ob}} \right) \\
&\quad - \operatorname{erf}\!\left( \frac{\vartheta_{ch}-\vartheta_{defl} + \sigma_{ob}^{2}/\vartheta_{dech}}{\sqrt{2}\,\sigma_{ob}} \right) \Bigg\},
\end{aligned}
\label{eq:dech}
\end{equation}
where \(A_{dech}\) is a normalization factor and $\vartheta_{dech}$ is the angular deviation subtended by $L_d$ along the curvature of the crystal. The fitting procedure was carried out in two steps. First, Eq.~\eqref{eq:hr} was applied to angular distributions in the amorphous orientation to determine the parameters $A_{ob}$ and $r$ (see Supplemental Material). These parameters were then kept fixed in the global fit of Eq.~\eqref{eq:total} to the complete distribution, which provided \(\vartheta_{dech}\) and thus \(L_d=R\vartheta_{dech}\) for both experimental and simulated distributions.

The resulting values are listed in Table~\ref{tab:L_dech} as \(L_{d,\text{exp fit}}\) and \(L_{d,\text{sim fit}}\). At 855 and 600 MeV, the values extracted from data and simulations agree within the uncertainties and follow the expected trend, decreasing from about 20 $\mu$m to about 15 $\mu$m. At 855 MeV, the dechanneling length is larger than the crystal thickness, so most channeled particles are expected to remain trapped over the full crystal length. At 600 MeV, $L_d$ becomes comparable to the thickness, and dechanneling starts to play a more significant role. For the 300 MeV case, the experimental angular distribution did not exhibit a well isolated channeling peak due to the increasing contribution of incoherent scattering.

To overcome the difficulties in extrapolating the dechanneling length from fits to the deflection distributions at lower energies and to explicitly quantify the role of rechanneling, we employed a different method, starting from  the simple exponential decay described in Eq.~\ref{eq:negexp}. In practice, we combined the initial capture efficiency $A_0$ together with the channeling population at depth $z$, $N_{ch}(z)$, obtained from Monte Carlo simulations, including the actual divergence of the beam.  The extraction was carried out for two simulation settings: one including rechanneling, yielding an effective dechanneling length $L_{d,\text{with rech}}$, which should be directly compared with the experimentally measured value, and one excluding rechanneling, providing the intrinsic dechanneling length $L_{d,\text{without rech}}$. 

The values of $L_{d,\text{with rech}}$ and $L_{d,\text{without rech}}$ are reported in table \ref{tab:L_dech} for all energies.
The values of $L_{d,\text{with rech}}$ are in good agreement with those obtained from the angular fit at 855 and 600 MeV. This method also enables an estimate at 300 MeV, yielding $L_{d,\text{with rech}} \approx 15~\mu\text{m}$. Although this confirms a more pronounced dechanneling effect at lower energies, the value is comparable to the crystal thickness, ensuring efficient steering also in this case. Finally, the comparison with $L_{d,\text{without rech}}$ highlights the substantial contribution of rechanneling, which significantly increases the dechanneling length by over 60\% at 300 MeV compared to the case in which rechanneling is absent.

\section{Radiation Generation}\label{subsecrad}

In this section, we present a detailed study of radiation emission by sub-GeV electrons interacting with the 15 $\mu$m bent silicon crystal described in Sec.~\ref{setup}. As in the Sec. \ref{sec2-Defl} which is devoted to beam steering, we investigate how the radiation emission varies as a function of beam energy and crystal orientation. Experimental spectra, collected using the NaI detector with an aperture of 2.32~mrad, are compared with Monte Carlo simulations performed using the \texttt{G4ChannelingFastSimModel}~\cite{Sytov-2024} described in Sec.~\ref{sec-simmethod}.

\begin{figure}
    \centering
    \includegraphics[width=1 \linewidth]{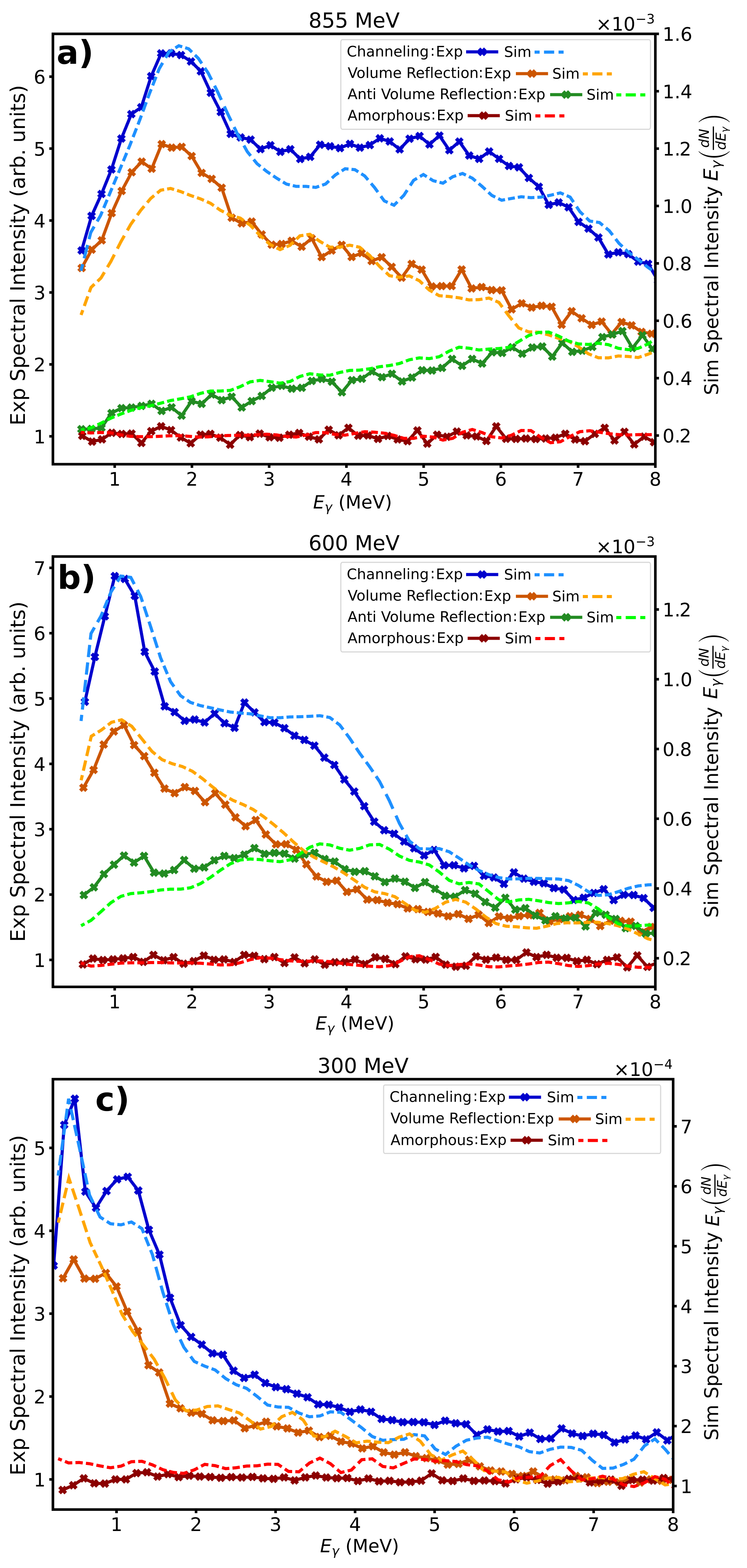}
    \caption{Experimental (solid lines with markers, left axis) and simulated (dashed lines, right axis) radiation spectral intensities for electron beam energies of 855~MeV (a), 600~MeV (b), and 300~MeV (c). Each orientation is associated with a specific color, using darker shades for experimental data and lighter shades for simulations: planar channeling (blue), volume reflection (orange), anti-volume reflection (green), and amorphous (red).}
    \label{fig:rad}
\end{figure}

Radiation spectra were collected for different crystal-to-beam orientations by scanning the crystal angle using the goniometer. The alignment angle $\alpha$ is defined relative to the optimal planar channeling condition, which is taken as the zero reference. Measurements were performed at four representative orientations: planar channeling ($\alpha \approx 0$) at all energies; amorphous (random orientation, $\alpha \gtrsim 10~\text{mrad}$) at all energies; volume reflection at $\alpha \approx 315~\mu\text{rad}$ for 855 and 600~MeV and $\alpha \approx 340~\mu\text{rad}$ for 300~MeV; and anti–volume reflection at $\alpha \approx -315~\mu\text{rad}$ for 855~MeV and $\alpha \approx -340~\mu\text{rad}$ for 600~MeV. Anti-volume reflection corresponds to an initial coherent bremsstrahlung condition, as is found at an angle slightly larger than $\vartheta_L$ but in the opposite direction with respect to volume reflection. 

Figure~\ref{fig:rad} presents the measured and simulated radiation spectral intensities for the three electron beam energies and for the different crystal-to-beam orientations under investigation. The experimental spectral intensities (solid lines, left axis) are obtained by multiplying the spectra collected by the NaI detector, which are proportional to the photon yield per unit time and energy, by the photon energy $E_{\gamma}$. The simulated spectral intensity corresponds to $E_{\gamma} \frac{dN}{dE_{\gamma}}$ (dashed lines, right axis), with $\frac{dN}{dE_{\gamma}}$ being the photon emission probability per electron calculated through the Baier--Katkov quasiclassical algorithm (see Sec.~\ref{sec-simmethod}). Since the experimental signal lacks an absolute normalization to the true photon count per electron, the two vertical scales are aligned by matching the simulated amorphous baseline to the experimental level. The comparison is therefore based on spectral shapes and relative enhancement factors with respect to the amorphous orientation. The photon energy range was deliberately restricted to 0.2--8~MeV to emphasize the regions most relevant for radiation enhancement in channeling and volume reflection at the investigated energies. Within this interval, radiation intensities are expected to be significantly higher than in the amorphous orientation~\cite{bandiera-2015,BACKE20083835}. Comparison of experimental and simulated data shows good agreement at 855 and 600~MeV for each investigated orientation, with the spectral shapes, peak positions, and relative enhancement factors well reproduced. At 300 MeV, the simulations are less effective in reproducing the experimental spectra: in the channeling orientation, the second harmonic peak is not fully reproduced, while in the volume reflection orientation, the agreement in the low-energy part of the spectrum deteriorates in terms of both spectral shape and relative intensity. These residual discrepancies are consistent with the stronger sensitivity to crystal imperfections and simulation approximations at lower energies.

\begin{table}[ht]
\centering

\fontsize{11}{14}\selectfont

\renewcommand{\arraystretch}{1}

\setlength{\tabcolsep}{8pt}

\caption{Radiation peak energy values for the three beam energies. All values are expressed in MeV.}
\label{tab:peaks}

\begin{tabular}{@{}cccc@{}}
\toprule
Energy & $E_{\text{peak}}^{\text{phenomena}}$ & $E_{\text{peak}}^{\text{exp}}$ & $E_{\text{peak}}^{\text{sim}}$ \\
\midrule
855 & 1.9 & 1.8 $\pm$ 0.2 & 1.9 $\pm$ 0.1 \\
600 & 1.2 & 1.1 $\pm$ 0.2 & 1.2 $\pm$ 0.1 \\
300 & 0.4 & 0.5 $\pm$ 0.1 & 0.4 $\pm$ 0.1 \\
\bottomrule
\end{tabular}

\end{table}

The channeling radiation spectra (blue curves) exhibit a pronounced first peak corresponding to the main channeling harmonic, while higher-order contributions populate the harder part of the spectrum. At sufficiently high photon energies, the distribution gradually approaches the amorphous level. The peak energy scales with the incident beam energy according to the analytical estimate obtained within the harmonic approximation of the planar potential~\cite{biryukov-1997}:
\begin{equation}
\hbar\omega =\frac{4\hbar c}{d_p} \sqrt{\frac{2 U_0}{mc^2}} \gamma^{3/2}
\label{eq:theory_peak}
\end{equation}where $d_p= 2.35 \AA$ and $U_0 \approx 21$ eV are the interplanar distance for (111) planes and the potential well depth, respectively; $m$ is the mass of the electron, and $\gamma$ is the Lorentz factor. The estimated values, summarized in Tab.~\ref{tab:peaks}, closely matched the peak positions obtained from Gaussian fits of the experimental data and simulations at different energies.

Radiation generated under volume reflection (orange curves) represents an intermediate regime between channeling radiation and coherent bremsstrahlung. Although its spectral intensity is lower than that of channeling radiation, it peaks at similar photon energies. This similarity arises from the contribution of volume captured particles, which undergo dynamics analogous to channeling. A key advantage of volume reflection radiation is its much larger angular acceptance compared to channeling, approximately equal to the crystal bending angle, making it a tunable radiation source for applications employing beams with large divergence.

The anti–volume reflection spectra (green curves) are characterized by a progressive increase in intensity toward the harder part of the spectrum, a behavior typical of coherent bremsstrahlung. The anti-volume reflection orientation at 300 MeV was not measured due to beamtime constraints, and is therefore not shown in Figure~\ref{fig:rad}(c).

Both channeling and volume reflection produce a substantial enhancement of radiation intensity relative to the amorphous baseline (red curves). At 855~MeV, the emission probability near the 1.8~MeV peak is enhanced by a factor of approximately 6.3 for channeling and 5.0 for volume reflection. At 600~MeV, the enhancement near the 1.1~MeV peak reaches factors of about 6.7 (channeling) and 4.6 (volume reflection). Even at 300~MeV, despite stronger multiple scattering, a robust enhancement persists at the 0.43~MeV peak, with factors of approximately 5.6 for channeling and 3.6 for volume reflection. The latter represents the first experimental demonstration that bent crystals can produce substantial radiation enhancement even at energies as low as 300~MeV. This result confirms the viability of bent crystals as compact radiation sources in an energy range covered by many small accelerators worldwide.

\subsubsection*{Dynamical Decomposition of Radiation Spectrum}

Analogous to the analysis performed for the beam deflection profiles 
in Sec.~\ref{sec2-Defl}, we applied the dynamical decomposition based 
on the evolution of the particles' transverse energy to the simulated 
radiation data. This allowed us to have a deep insight into the specific 
contribution of each motion regime to the total spectral intensity in 
the channeling orientation~\cite{negrello-2025}.

\begin{figure}
    \centering
    \includegraphics[width=0.97\linewidth]{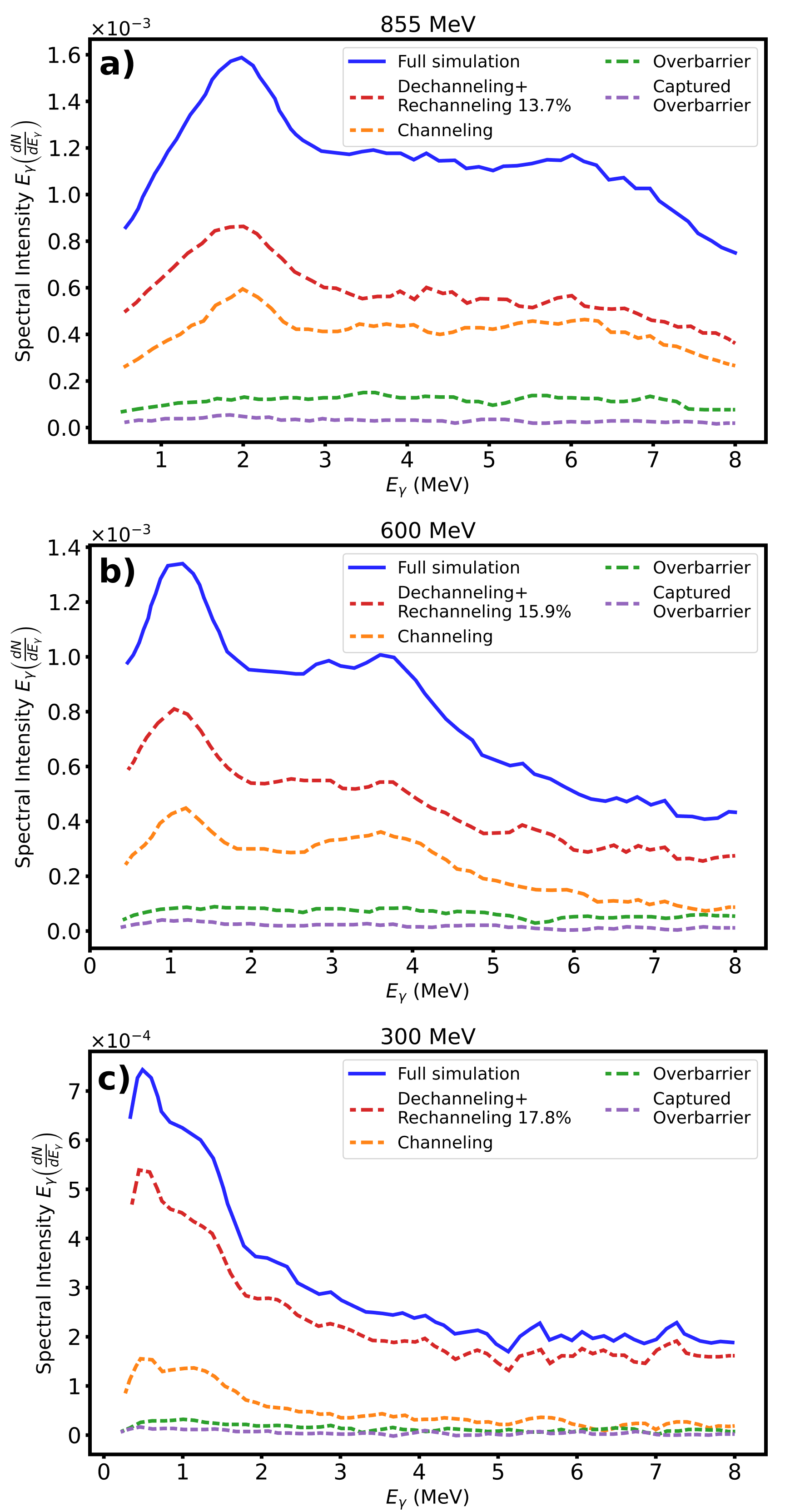}
    \caption{Simulated spectral intensity at (a) 855~MeV, (b) 600~MeV, 
    and (c) 300~MeV: total spectrum (blue) and contributions from stable 
    channeling (orange), dechanneling + rechanneling (red), overbarrier 
    (green), and captured overbarrier (purple).}
    \label{fig:tagrad}
\end{figure}

Figure~\ref{fig:tagrad} presents the decomposition of the spectral 
intensity for the channeling orientation across the three beam energies. 
The solid blue line shows the full simulation, while dashed lines 
represent the contributions from the dynamical categories of particles: 
stable channeling (orange), dechanneling+rechanneling (red), overbarrier 
motion (green), and captured overbarrier (purple), as defined in the 
beam steering analysis (Sec.~\ref{sec2-Defl}). In this representation, 
the contribution of dechanneled and rechanneled has been grouped 
together, as these two regimes are dynamically linked through multiple 
transitions.

To characterize these trends quantitatively, 
Table~\ref{tab:efficiency_summary} reports the contribution to the 
total yield for each dynamical class, computed as the fraction of the 
total spectrum. To gain further insight into the total yield, 
Table~\ref{tab:efficiency_summary} distinguishes the contribution of 
rechanneled particles from the purely dechanneled ones.

\begin{table*}[ht]
\centering
\caption{Contribution to the total photon yield and radiation 
efficiency per particle for different dynamical categories and beam 
energies, evaluated in the photon energy range [0, 8]~MeV.}
\label{tab:efficiency_summary}
\begin{tabular}{@{}cccc@{}}
\toprule
Beam Energy [MeV] & Category & Contribution to yield (\%) & 
Radiation Efficiency ($10^{-3}$) \\
\midrule
\multirow{5}{*}{855} 
& Channeled             & 34.1 & 4.9 \\
& Overbarrier          & 9.3  & 1.5 \\
& Dechanneled           & 46.3 & 3.8 \\
& Rechanneled           & 7.4  & 4.7 \\
& Captured Overbarrier & 2.9  & 4.7 \\
\midrule
\multirow{5}{*}{600} 
& Channeled             & 30.5 & 4.2 \\
& Overbarrier          & 7.3  & 1.4 \\
& Dechanneled           & 50.0 & 3.1 \\
& Rechanneled           & 9.4  & 3.9 \\
& Captured Overbarrier & 2.8  & 3.6 \\
\midrule
\multirow{5}{*}{300} 
& Channeled             & 19.6 & 2.3 \\
& Overbarrier          & 4.2  & 0.9 \\
& Dechanneled           & 60.9 & 1.8 \\
& Rechanneled           & 13.1 & 1.8 \\
& Captured Overbarrier & 2.2  & 1.8 \\
\bottomrule
\end{tabular}
\end{table*}

In particular, dechanneled particles provide the largest contribution 
to the photon yield at all investigated energies, increasing from 
$46.3\%$ at 855~MeV to $60.9\%$ at 300~MeV 
(Table~\ref{tab:efficiency_summary}). This increase should not be 
interpreted as a reduced role of channeling radiation, but rather as a 
consequence of enhanced multiple scattering at lower energies, which 
transfers particles from stable channeling to trajectory histories 
involving dechanneling and possible recapture. Accordingly, the 
contribution of stably channeled particles decreases from $34.1\%$ to 
$19.6\%$. Since dechanneled and rechanneled particles spend part of 
their trajectories undergoing channeling oscillations around the atomic 
planes, their spectral profiles retain the main features associated with 
channeling radiation. The purely overbarrier fraction also diminishes 
with decreasing energy (from $9.3\%$ to $4.2\%$) as more particles 
enter bound states thanks to the increasing $\vartheta_L^b$ at lower 
energies. Interestingly, the rechanneling contribution nearly doubles 
(from $7.4\%$ to $13.1\%$) as energy decreases, suggesting that 
particles undergo more frequent transitions between overbarrier and 
channeling states at lower energies. The lower contribution of stably 
channeled particles at 300~MeV leads to a smaller enhancement factor 
of channeling radiation with respect to the amorphous orientation 
(5.6 at 300~MeV versus 6.3 at 855~MeV).

Tab.~\ref{tab:efficiency_summary} also reports the per-particle 
radiation efficiency of each category, defined as 
the fraction of particles in the class emitting at least one photon. 
Stably channeled particles (class ``Channeled'') show the highest 
per-particle efficiency at every energy, closely followed by rechanneled 
and captured-overbarrier particles, confirming that the emission 
probability is governed by the fraction of the trajectory spent in 
channeling. In contrast, particles classified as subjected to 
dechanneling (and not being rechanneled), despite providing the bulk of 
the photons due to their high population, exhibit moderate radiation 
efficiency per particle, which decreases at lower energy, indicating 
that while many particles enter this regime, the probability of 
effective radiation per particle is not maximized. The captured 
overbarrier regime also shows relatively high efficiencies, suggesting 
that once capture occurs, the resulting emission probability resembles 
that of regular channeling, despite its low population fraction. As 
expected, particles undergoing overbarrier motion show consistently 
lower efficiencies than any other category.

Across all beam energies, the contributions involving particles in 
channeling, whether directly or via rechanneling or capture, show a 
marked increase near the spectral peak. This underlines the central role 
of channeling oscillations in shaping the peak structure of the emitted 
radiation.

To further interpret these values and quantify the degree of 
optimization of the crystal parameters, we adopt an exponential 
saturation model for the channeling-radiation efficiency,
\begin{equation}
\varepsilon_{ch}(\ell)
=  \varepsilon^{\max}_{ch}\!\left(1 - e^{-\ell/L_d}\right),
\end{equation}
where $\varepsilon_{ch}(\ell)$ denotes the channeling-radiation 
efficiency for a crystal of thickness $\ell$, $\varepsilon^{\max}_{ch}$ 
is the asymptotic efficiency reached for $\ell \gg L_d$, and $L_d$ is 
the dechanneling length.

Using the simulated channeling efficiencies reported in 
Tab.~\ref{tab:efficiency_summary} for a crystal thickness of 
$\ell = 15~\mu\mathrm{m}$ (i.e., $4.9\times10^{-3}$ at 855~MeV, 
$4.2\times10^{-3}$ at 600~MeV, and $2.3\times10^{-3}$ at 300~MeV), 
together with the corresponding dechanneling lengths obtained from 
simulation and listed in Tab.~\ref{tab:L_dech} ($L_{d,\text{with 
rech}}$), we infer asymptotic efficiencies of $\varepsilon^{\max}_{ch} 
= 9.6\times10^{-3}$, $7.4\times10^{-3}$, and $3.6\times10^{-3}$.

For the investigated thickness $\ell = 15~\mu\mathrm{m}$, this implies 
that only about $51\%$ (855~MeV), $57\%$ (600~MeV), and $63\%$ 
(300~MeV) of the asymptotic channeling-radiation yield is achieved. As 
a practical benchmark for hard X-ray and $\gamma$-ray source 
applications, collecting $90\%$ of the asymptotic yield would require a 
crystal thickness of approximately $\ell \simeq 2.3\,L_d$.

Although simplified, this effective description provides a practical 
tool to guide the preliminary optimization of crystal thickness for 
channeling-based radiation sources. A full optimization of the crystal geometry would, 
however, need to account also for other secondary effects shown by the 
present results to be quantitatively significant, such as rechanneling 
(Tabs.~\ref{tab:L_dech} and~\ref{tab:efficiency_summary}) and photon 
self-absorption in thicker crystals. Such an extension is left for 
future work.

\section{Discussion and Conclusions}\label{sec5-discussion}

This study provided a combined experimental and numerical investigation of beam steering and radiation generation by sub-GeV electrons interacting with a 15~$\mu$m-long bent silicon crystal. The study explored the energy range of 300--855~MeV, providing a first experimental demonstration of enhanced radiation emission from a bent crystal at 300~MeV.

A central outcome is the joint analysis of deflection profiles and radiation spectra as a function of beam energy and crystal orientation. Experimental measurements were compared with Monte Carlo simulations performed with the new Geant4-based model~\cite{Sytov-2024}, showing good overall agreement in terms of spectral shape, peak positions (Tab.~\ref{tab:peaks}), and relative enhancement factors with respect to the amorphous orientation. By applying trajectory classification to the simulated events, we examine the links between steering regimes (e.g.\ channeling, volume reflection, dechanneling, and rechanneling) and their radiation signatures, showing how the microscopic dynamics influences both spectral features and the overall emission efficiency.

The deflection measurements show that bent silicon crystals remain highly effective down to the lowest investigated energy. At 300~MeV, we measured a channeling efficiency of 54\%. This performance was enabled by the fabrication of ultrathin crystals, with a thickness of the order of the dechanneling length, which mitigates the impact of multiple scattering and preserves a high channeling probability. This value is more than five times higher than previous results at comparable energies (e.g., those obtained at SAGA~\cite{TAKABAYASHI2018153}) and sets a new benchmark in this regime. The increase in efficiency at lower energy can be attributed to the favourable ratio between the crystal bending radius and the critical radius ($R/R_C$), which increases the phase-space acceptance for channeling. Trajectory classification further suggests that rechanneling becomes particularly relevant at 300~MeV, partially compensating for stronger multiple scattering and a shorter dechanneling length, thereby helping to sustain the deflected peak.

Regarding radiation, both the measured and simulated spectra show a clear enhancement with respect to random orientation. In channeling, a pronounced spectral peak is observed, shifting from about 0.5~MeV at 300~MeV to about 2~MeV at 855~MeV; its position is consistently reproduced by modeling, simulations and experiments across all investigated energies. Volume-reflection spectra exhibit a peak at comparable photon energies; although weaker than in channeling, the much larger angular acceptance may be advantageous for intense radiation generation when beams with large divergence are employed. The enhancement with respect to standard bremsstrahlung decreases with increasing beam energy. Nevertheless, even at 300~MeV the measured enhancement relative to random orientation reaches about 5.6 in channeling and about 3.6 in volume reflection. As a practical estimate, a 10 $\mu$A, 300 MeV electron beam traversing the 15 $\mu$m bent crystal in channeling orientation would produce approximately $3 \times 10^{10}$ photons/s in the channeling peak region (0.3-0.6 MeV), about five times higher than the bremsstrahlung yield from the same crystal in amorphous orientation. For channeling radiation, simulations indicate that the peak originates from both stable channeling and a substantial contribution from dechanneled and rechanneled electrons.

Many accelerator facilities operate with electron beams in the few-hundred-MeV range and could benefit from the use of bent crystals. 
First, our steering results indicate that bent crystals may be considered for beam steering, collimation, and slow extraction in sub-GeV accelerators such as MAMI, SAGA (up to 255~MeV)~\cite{TAKABAYASHI2018153}, CLARA (up to 250~MeV)~\cite{Snedden2024}, and CLEAR@CERN (up to 230~MeV)~\cite{Corsini2022}, as well as at lower-energy facilities such as ARES@DESY (up to 160~MeV)~\cite{ARES2024} and SPARC\_LAB@INFN--LNF (up to 180~MeV)~\cite{SPARCLAB}. In compact layouts where space constraints limit conventional magnetic solutions, the measured deflection efficiencies, together with the larger critical angle at lower energies, can relax alignment requirements and enlarge operational margins. This may be particularly relevant for collimation and extraction schemes, as well as for diagnostic beamlines.

In addition, the observed radiation enhancement with respect to standard bremsstrahlung supports the feasibility of generating enhanced hard X- and $\gamma$-rays using bent crystals. Photon-yield enhancements may remain achievable at lower energies and for larger beam divergence, given the approximate inverse scaling of channeling acceptance with energy. For instance, at 100~MeV the expected channeling peak in the 60--80~keV range could be of interest for X-ray microscopy~\cite{VillanuevaPerez2021}. More broadly, these findings may be relevant for applications such as industrial radiography (e.g., Non-Destructive Testing), where photon energies in the 80--340~keV range are commonly employed~\cite{Adeli2022}. Bent crystals offer a route to tune the radiation output via the beam energy and to exploit volume reflection, which is typically less sensitive to crystal defects than channeling and can operate with lower-quality beams and high-$Z$ materials such as W~\cite{posiprod_w_tungsten, gao-2023}. This may broaden the range of practical use cases for crystalline radiation sources~\cite{Bagli2014effVsRad, bandiera-2015}.

The present results are also directly relevant to the design and optimization of advanced radiation sources based on periodically bent crystals (crystalline undulators)~\cite{baryshevsky1980generation, korol1998coherent, korol-2000, baryshevsky2013crystal, kostyuk2014realization, korol-2014, korol2020crystal, pavlov-2020, korol2022novel}, as well as to systems exploiting multi-volume reflection~\cite{tikhomirov-2007, guidi2010, guidi2012radiation, bandiera2013multiVR}.

\section*{Funding}
\small

\noindent This work was supported by the European Commission through the H2020-MSCA-RISE N-LIGHT (Grant Agreement No. 872196), EIC-PATHFINDER-OPEN TECHNO-CLS (Grant Agreement No. 101046458), and INFN CSN5 (CORAL and GEANT4-INFN projects). \\
We acknowledge partial support from the European Union -- NextGenerationEU under the project ``Intense positron source Based On Oriented crySTals - e+BOOST'' (Grant No. 2022Y87K7X, CUP I53D23001510006).

\newpage
\bibliographystyle{unsrt}
\bibliography{sn-bibliography}

\end{document}